\documentclass[twocolumn,aps,10pt,showpacs,showkeys,]{revtex4}
\usepackage[dvips]{graphicx}

\begin{document}
\title{Comment on a Phys. Rev. Lett. paper: "All-Electron Self-Consistent GW Approximation: Application to Si, MnO, and NiO". Magnetic moment of NiO}
\author{R. J. Radwanski}
\affiliation{Center of Solid State Physics, S$^{nt}$Filip 5,
31-150 Krakow, Poland,
\\
Institute of Physics, Pedagogical University, 30-084 Krakow,
Poland} \homepage{http://www.css-physics.edu.pl}
\email{sfradwan@cyf-kr.edu.pl}
\author{Z. Ropka}
\affiliation{Center of Solid State Physics, S$^{nt}$Filip 5,
31-150 Krakow, Poland}

\begin{abstract}
We claim that any approach neglecting the spin-orbit coupling and
the orbital magnetism is not physically adequate for 3d oxides,
including NiO, and that in reaching "excellent agreement" in a
Phys. Rev. Lett. {\bf 93}, 126406 (2004) paper too small
experimental value of 1.9 $\mu _{B}$ has been taken for the Ni
magnetic moment despite publication of a new experimental value
of 2.2 $\mu _{B}$, at 300 K yielding 2.6 $\mu _{B}$ at T = 0 K,
already in a year of 1998.

\end{abstract}
\maketitle

By this Comment we would like to express our deep scepticism about
"Excellent agreement with experiment for many properties" of NiO
claimed in the abstract of a paper in Phys. Rev. Lett. {\bf 93},
126406 (2004) by Faleev et al. \cite {1} which has been obtained
with "a new kind of self-consistent GW (SCGW) approximation based
on the all-electron, full potential linear muffin-tin orbital
method."

This excellent agreement in the SCGW approach is based, among
others but this can be verified, on the obtained value of 1.72
$\mu _{B}$ for the Ni magnetic moment. This value is indeed
improved, in comparison to 1.28 $\mu _{B}$ obtained within the
LDA, becoming closer to an experimental value of 1.9 $\mu _{B}$.
However, we claim that this experimental value is presently
documented to be wrong, is too small, though this value has often
been quoted in literature in last 30 years.

The literature for the 1.9 $\mu _{B}$ value has not been given in
the commented paper \cite{1} but likely it is quoted after Ref.
\cite{2} from 1983. Earlier experimental papers have provided 1.64
$\mu _{B}$ \cite{3} in 1962 and 1.77 $\mu _{B}$ \cite{4} in 1968.
Here we would like to put attention that the recent very-detailed
experiment of the Grenoble group from 1998 has provided 2.2$\pm
$0.3 $\mu _{B}$ at 300 K \cite{5}. This experimental finding has
been recalled already in Refs \cite {6,7,8,9}. We made use of
this value in discussion of our calculations for the spin and
orbital moment in NiO in a Letter to Phys. Rev. Lett. (LF7313,
submitted 8.06.1999) \cite{8}. Our point of the Comment is that
even accepting the agreement between the calculated value of 1.72
$\mu _{B}$ and experiment of 1.9 $\mu _{B}$, but never excellent,
the calculated value 1.72 $\mu _{B}$ is substantially too small
with respect to the real value in NiO. The disagreement becomes
larger if one realizes that the experimental value of 2.2 $\mu
_{B}$ was derived at 300 K. Extrapolation of this value to the
zero temperature by means of a well-known equation m(T)/m(0)=
[(1-($T/T_{N}$)$^{2}$]$^{1/2}$ leads to 2.6 $\mu _{B}$ at T = 0
K. In such circumstances the claimed "excellent agreement" is not
at all justified. The acceptance by Editors and referees of Phys.
Rev. Lett. of this excellent agreement (10$\%$, what should we
write in case of 1-2$\%$ agreement???) would indicate their
conviction that the understanding of the magnetism and electronic
structure of NiO has been finally solved or at least largely
improved. By expressing our scepticism we would like to say that
it is not at all the case. It is a bad overlook of the Editor and
referees allowing for the neglect of all works of the Grenoble
group on NiO. It is interesting that this overlook in Phys. Rev.
Lett. coincides with the discrimination of our papers which for
justification of our atomic-like calculations for NiO recall this
larger moment value - this problem goes, however, beyond the
present Comment. Surely this discrimination has contributed to
weak spreading of the novel value of the magnetic moment of NiO.

In this place, being discriminated in Phys. Rev. Lett., we would
like to put attention to our atomistic approach to NiO based on
an assumption that the paramagnetic atom/ion preserves largely
its integrity also in a solid \cite{10,11}. We start analysis of
NiO from the detailed analysis of the single-ion effects like the
low-energy electronic structure of Ni$^{2+}$ ion \cite{8,9}.
Within the Quantum Atomistic Solid State theory (QUASST), taking
into account strong electron correlations, basically of the
on-site origin, the intraatomic spin-orbit coupling,
crystal-field interactions, yielding discrete energy states,
completed with inter-site spin-dependent interactions we have
calculated the Ni magnetic moment at T = 0 K as 2.54 $\mu _{B}$,
i.e. really very close to the experimental value. Moreover, we
have calculated the orbital and spin contributions to the
magnetic moment as well as physically adequate thermodynamics
\cite{12}. We describe in the consistent way both the paramagnetic
and the magnetic state with the description of, for instance, the
$\lambda$-peak at $T_{N}$ in the temperature dependence of the
heat capacity.

In conclusion, we claim that any approach neglecting the
spin-orbit coupling and the orbital magnetism is not physically
adequate to 3d oxides and that in reaching "excellent agreement"
too small experimental value of 1.9 $\mu _{B}$ instead of 2.2
$\mu _{B}$, or even 2.6 $\mu _{B}$, has been taken for the Ni
magnetic moment in NiO.

\end{document}